\documentclass[preprint,showpacs,preprintnumbers,amsmath,amssymb]{revtex4}

\usepackage{graphicx}% Include figure files
\usepackage{dcolumn}% Align table columns on decimal point
\usepackage{bm}% bold math

\begin{document}

%\preprint{APS/BNK1011_deleo}

\title{Comment on ``Thermal propagation in two-dimensional Josephson junction arrays''}

\author{Cinzia De Leo}
\affiliation{
 Dipartimento di Ingegneria Meccanica, Energetica e Gestionale \\ University of L'Aquila, L'Aquila, I-67040 Italy }%
\email{deleo@ing.univaq.it}

\date{\today}% It is always \today, today,

\begin{abstract}
In a recent paper, Filatrella et al. [Phys. Rev. B 75, 54510 (2007)] report results of numerical calculations of energy barriers for flux quanta propagation in two-dimensional arrays of Josephson junctions with finite self and mutual inductances. To avoid complex numerical calculations, they use an approximated inductance model to address the effects of the mutual couplings. Using a full inductance matrix model, we show that this approximated model cannot be used to calculate the energy barrier of interacting vortices.
The authors' method of comparison between numerical data and theoretical predictions is also discussed.
\end{abstract}

\maketitle

The static properties of single-vortex states in two-dimensional arrays of Josephson junctions have been analyzed by Phillips et al.\cite{Phillips} when the self field effects are important. Using numerical calculations, they found that the full inductance matrix is necessary to recall the magnetic properties of the vortex that are similar to the magnetic properties of a vortex in a thin superconducting film.They also showed that it is sufficient to include self and nearest- neighbour terms in the inductance matrix to model the energy barrier for the cell-to-cell motion of an isolated vortex. In fact depending on the structure of the vortex near the core, the energy barrier is less affected by the mutual inductances, while the vortex properties that are related to longer length scale, such as the vortex shape or the current distribution, are more strongly affected. Moreover, they stressed that when a vortex model includes only the self terms of the inductance matrix, the magnetic properties of a vortex are similar to those of a vortex in a bulk superconductor. This vortex model is sometimes referred to as the Nakajima-Sawada model \cite{Nakajima}.

In a recent paper Filatrella et al.\cite{Filatrella} present a numerical study on the effect of noise on vortex propagation in two-dimensional arrays of Josephson junctions. They e\-va\-lu\-a\-te the energy barrier in the vortex-(anti)vortex interaction. Two models of vortex are considered: the Nakajima-Sawada (NS) model and the mutual coupling (MC) model, which is an approximated inductance model, developed by the authors, to address the effects of the mutual couplings between the array elements and thus avoid the complex numerical calculations of a full inductance matrix model \cite{Rotoli}. They calculate the energy barriers for the cell-to-cell motion of an isolated vortex to test the goodness of the MC model. The compatibility of the findings with the numerical results of Phillips et al.\cite{Phillips} is considered by the authors as a good test for the MC model.

We observe that although a simple self and nearest-neighbour inductance model is sufficient for consistency with the results of Phillips et al.\cite{Phillips} it is necessary to demonstrate that an approximate inductance model, such as the MC model, is equivalent to a fully inductance model \cite{pc}.

Figure~5(d) of Filatrella et al.\cite{Filatrella} shows the energy barrier for the cell-to-cell vortex motion as a function of the distance between two vortices. This is the case in which the energy barrier adds the contribution of the vortex-vortex interaction. The comparison between the data obtained from the NS model with those from the MC model predicts  that the energy barrier of two interacting vortices in a bulk superconductor cannot be distinguished from that of two interacting vortices in a thin superconducting film. In other words, the characteristic long-range interaction of two vortices in a thin film \cite{Pearl} does not produce a clear signal for the vortex motion.

To show that this result depends on the method developed by the authors to address the mutual magnetic couplings \cite{Filatrella}-\cite{Rotoli}, we have repeated the analysis using a full inductance matrix model \cite{DeLeo} .

In Figure~\ref{fig:barrier} we show the comparison between the NS model with the full inductance model at $\beta_L = 0.5$. The data clearly show that the energy barrier of two interacting vortices in a full inductance matrix model is different from that of two interacting vortices in a model that includes only the self terms of the inductance matrix. Unlike Filatrella et al.\cite{Filatrella}, we find that the mi\-ni\-mum distance for a static pair of vortices also depends on the model, as consequence of the different behaviour of the interaction. This distance is equal to three cells for the NS model and to four cells for the full inductance matrix model.

 Moreover, from another point of view, the force of interaction between two vortices depends on the vortex current produced by the first vortex at the centre of the second vortex \cite{shmidt} and, as  Phillips et al.\cite{Phillips} have shown, the correct calculation of the current distribution requires to include all terms in the mutual inductance matrix. Therefore, an approximated inductance model is not able to simulate  the long-range interaction between vortices.

The second consideration on the study presented by Filatrella et al.\cite{Filatrella} is the com\-pa\-ri\-son between numerical data and theoretical predictions.

For a single vortex-state the energy barrier height $\Delta \epsilon_0$ is given by the energy difference  between a vortex centered on a junction and a vortex centered in the middle of the cell to which the junction belongs \cite{Lobb}.

 For two interacting vortices at a relative cell distance $d$, the energy barrier height is calculated, by the authors, as the energy necessary to shift one of the vortices by half a cell  \cite{Rotoli} and  it is splitted in two terms:
\begin{eqnarray}
\Delta \epsilon = \Delta \epsilon_0 + \Delta \epsilon^{int}
\label{eq:preo}
\end{eqnarray}
where $\Delta \epsilon_0$ is the term due to the vortex interaction with the network and $\Delta \epsilon^{int}$ the term due to the interaction between vortices, connected to the vortex-vortex interaction potential $\epsilon^{int}$.
In particular, regarding the MC model, Filatrella et al.\cite{Filatrella} analyze the behaviour of $\Delta \epsilon^{int}$ by fitting the numerical data with a power law as a function of the distance $ d$ between the vortices:
\begin{eqnarray}
-\Delta \epsilon^{int} = \frac{A}{d^\alpha}
\end{eqnarray}
and find $A \simeq 3$, in units of Josephson energy, and $\alpha \simeq 3.1$. The last value is compared with the theoretical prediction $ \alpha^{th}=1$ obtained by  Dominguez and  Jos\'e \cite{Dominguez} for the vortex-vortex interaction energy of a full inductance model.

We observe that this comparison is not correct because the two quantities are not phy\-si\-cal\-ly equivalent, since the prediction by Dominguez and Jos\'e \cite{Dominguez} refers to the interaction energy, i.e. the energy necessary to put two vortices, starting from infinity, at a finite distance $d$:
\begin{eqnarray}
\Delta \epsilon^{int} = \epsilon^{int}(d)- \epsilon^{int}(\infty)= \frac{A}{d}
\end{eqnarray}
while $\Delta \epsilon^{int}$ for  Filatrella et al.\cite{Filatrella} is the contribution of the vortex-vortex interaction to the energy barrier, i.e. the change of the interaction potential when, starting with two vortices at a finite distance $d$, one of them is displaced, toward $(d \rightarrow d-1/2)$ or apart $(d \rightarrow d+1/2)$, by half a cell through the lattice barrier. Therefore, the right predictions for the two cases are:
\begin{eqnarray}
-\Delta \epsilon^{int} = \epsilon^{int}(d)- \epsilon^{int}(d- 1/2)={\frac{A}{d}-\frac{A}{d-1/2}}  \qquad     d \rightarrow d-1/2
\\
-\Delta \epsilon^{int} = \epsilon^{int}(d)- \epsilon^{int}(d+ 1/2)={\frac{A}{d}-\frac{A}{d + 1/2}}   \qquad   d \rightarrow d+1/2 \\
\label{eq:pre}
\end{eqnarray}
Using a full inductance matrix model \cite{DeLeo} we have analyzed the behaviour of $\Delta \epsilon^{int}$ when one vortex is moved apart by half a cell in a $75\times 75$ square array at $\beta_L = 0.5 $. Fitting the data reported in Figure~\ref{fig:interaction} with the function:
\begin{eqnarray}
-\Delta \epsilon^{int} = \frac{A}{d^\alpha}-\frac{A}{d^\alpha + 1/2}
\label{eq:alpha}
\end{eqnarray}
we find A=7.87 units of Josephson energy and $\alpha =0.85 $. The result is in good agreement with the theoretical prediction by Dominguez and Jos\'e \cite{Dominguez}.\\

This work has been partially supported by MIUR PRIN 2006 under
the project: "Macroscopic Quantum Systems - Fundamental
Aspects and Applications of Non-conventional Josephson Structures"


\newpage
\maketitle{FIGURE CAPTIONS}
\\
\\

%\section*{FIGURE CAPTIONS}

\begin{figure}[h]
\caption{\label{fig:barrier} Energy barriers for two vortices in units of Josephson energy as a function of the distance between the vortices. Comparison of the NS model (filled symbols) with the full inductance matrix model (empty symbols) for $\beta_L=0.5$ in the case of vortex-vortex (circles) and vortex-antivortex (triangles) interactions.}
\end{figure}

\begin{figure}[h]
\caption{\label{fig:interaction} Normalized interaction energy contributes to the pinning barriers for two vortices. Vortex-vortex pair in a full inductance matrix model with $\beta_L = 0.5$. The solid line has been obtained fitting the data with the function of Eq.~\ref{eq:alpha}}
\end{figure}
\newpage
\begin{figure*}[h]
\includegraphics[width=17.8cm]{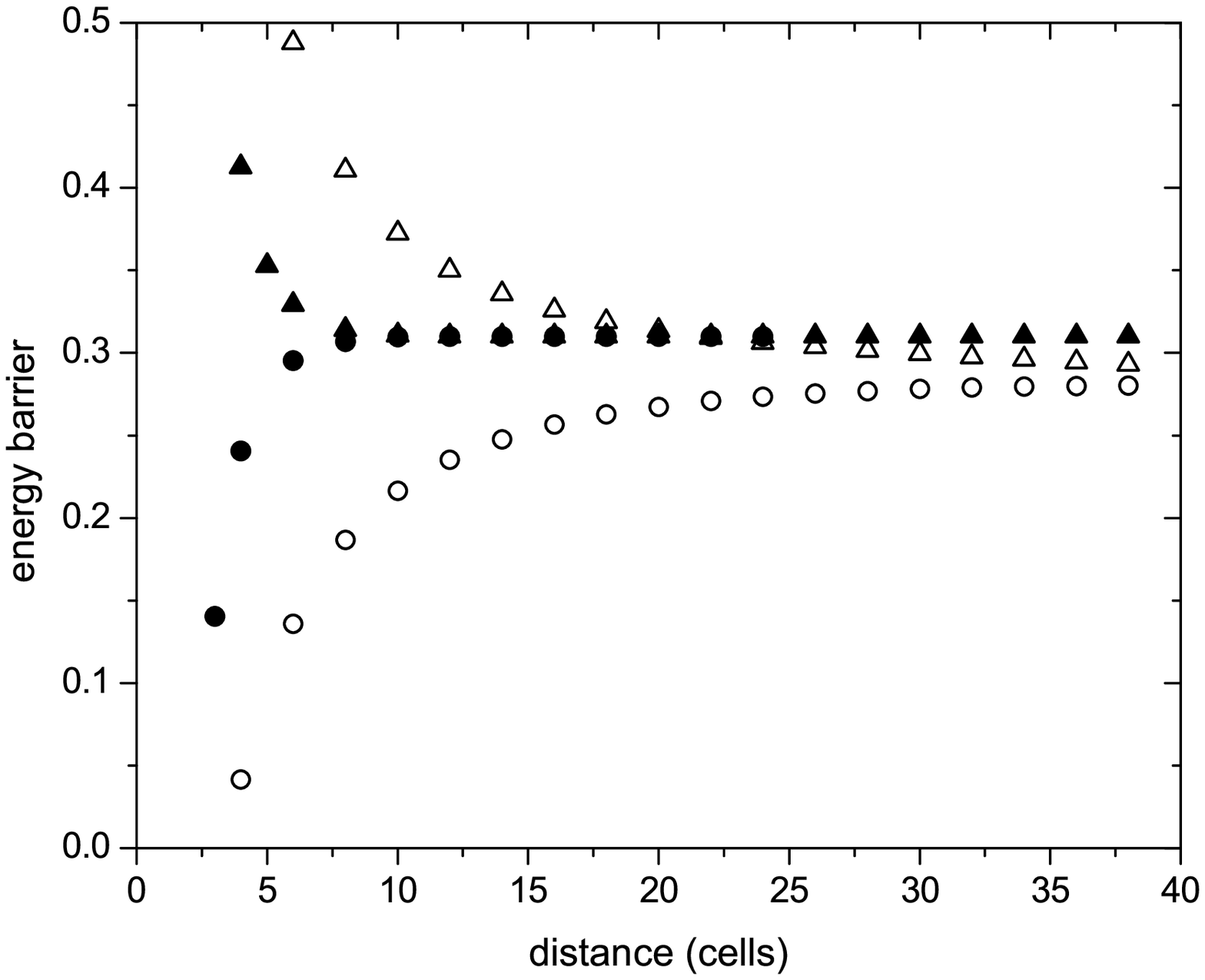}
\end{figure*}
\newpage
\begin{figure*}[h]
\includegraphics[width=17.8cm]{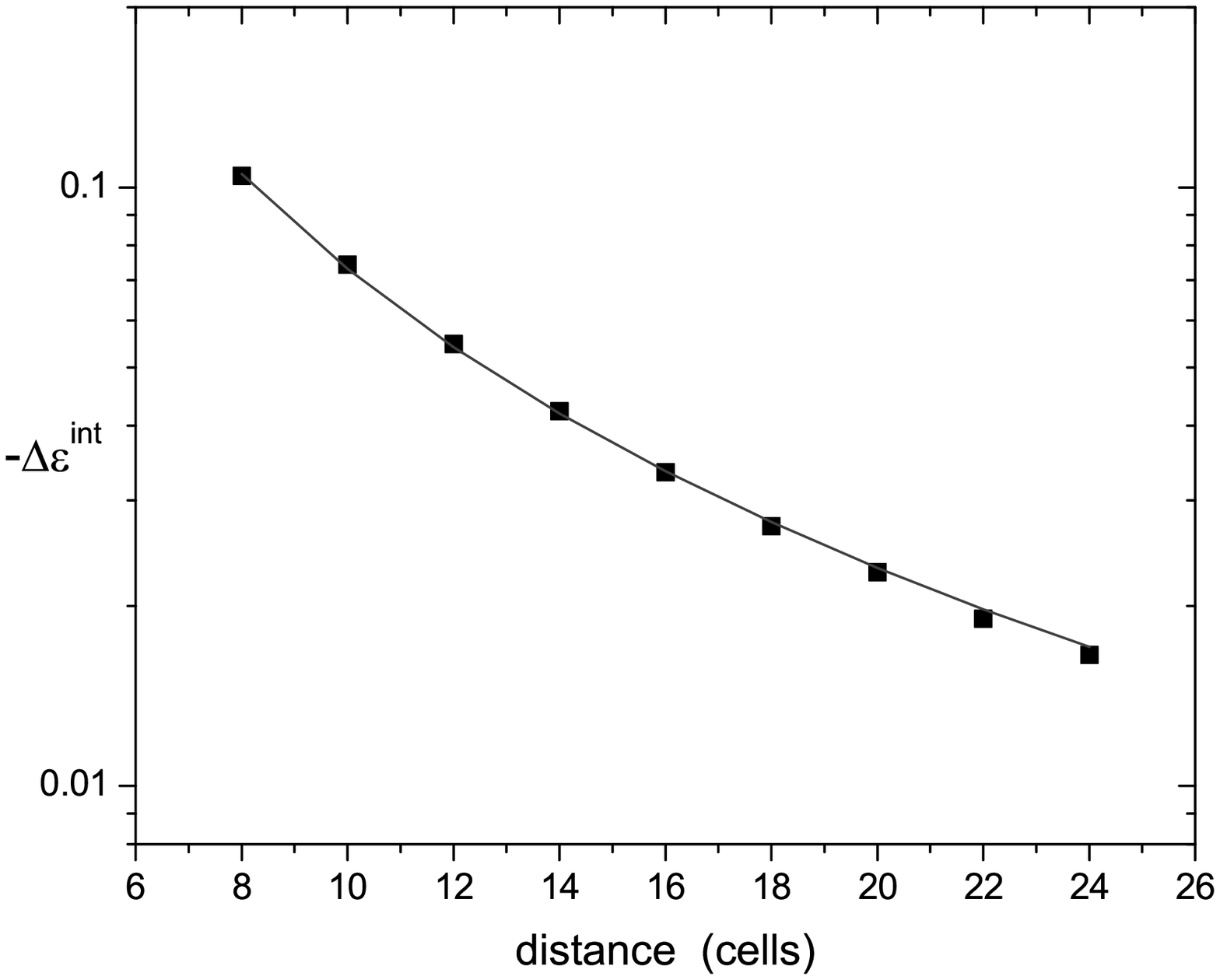}
\end{figure*}

\
\end{document}